\documentclass[]{elsart}

\usepackage{amsfonts}
\usepackage{amsmath}
\usepackage{amssymb}
\usepackage{graphicx}

\journal{JOURNAL OF ALLOYS AND COMPOUNDS}
\begin{document}
\begin{frontmatter}



\title{Valence fluctuation in CeMo$_{2}$Si$_{2}$C}


\author[iit]{U. B. Paramanik},
\author[iit]{Anupam},
\author[mp]{U. Burkhard},
\author[iit]{R. Prasad},
\author[mp]{C. Geibel},
\author[cor1,iit]{Z. Hossain},
\corauth[cor1]{Corresponding author: Tel.: +91 51 2679 7464.
E-mail address: zakir@iitk.ac.in}
\address[iit]{Department of Physics, Indian Institute of Technology, Kanpur 208016, India}
\address[mp]{Max-Planck Institute for Chemical Physics of Solids, 01187 Dresden, Germany}
\begin{abstract}
We report on the valence fluctuation of Ce in CeMo$_{2}$Si$_{2}$C as studied by means of magnetic susceptibility $\chi(T)$, specific heat $C(T)$, electrical resistivity $\rho(T)$ and x-ray absorption spectroscopy. Powder x-ray diffraction revealed that CeMo$_{2}$Si$_{2}$C crystallizes in CeCr$_{2}$Si$_{2}$C-type layered tetragonal crystal structure (space group \textit{P4/mmm}). The unit cell volume of CeMo$_{2}$Si$_{2}$C deviates from the expected lanthanide contraction, indicating non-trivalent state of Ce ions in this compound. The observed weak temperature dependence of the magnetic susceptibility and its low value indicate that Ce ions are in valence fluctuating state. The formal $L_{III}$ Ce valence in CeMo$_{2}$Si$_{2}$C  $<$$\widetilde{\nu}$$>$ = 3.11 as determined from x-ray absorption spectroscopy measurement is well bellow the value $<$$\widetilde{\nu}$$>  \simeq$ 3.4 in tetravalent Ce compound CeO$_{2}$. The temperature dependence of specific heat does not show any anomaly down to 1.8~K which rules out any magnetic ordering in the system. The Sommerfeld coefficient obtained from the specific heat data is $\gamma$ = 23.4~mJ/mol\,K$^{2}$. The electrical resistivity follows the $T{^2}$ behavior in the low temperature range below 35~K confirming a Fermi liquid behavior. Accordingly both the Kadowaki Wood ratio $A/\gamma^{2}$ and the Sommerfeld Wilson ratio $\chi(0)/\gamma$ are in the range expected for Fermi-liquid systems. In order to get some information on the electronic states, we calculated the band structure within the density functional theory, eventhough this approach is not able to treat $4f$ electrons accurately. The non-$f$ electron states crossing the Fermi level have mostly Mo $4d$ character. They provide the states with which the $4f$ sates are strongly hybridized, leading to the intermediate valent state.
\end{abstract}

\end{frontmatter}

\section{Introduction}
\label{sec:Intro}
The Ce-based intermetallic compounds have attracted tremendous attention with interesting phenomena such as valence fluctuations, heavy fermion behavior, quantum criticality, unconventional superconductivity and Kondo effect, etc.\cite{Wohlleben, Stewart, Brandt, Gegenwart} These anomalous properties arise due to the hybridization between the localized $4f$ electrons and conduction electrons. Usually, the competition between Ruderman-Kittel-Kasuya-Yosida (RKKY) and Kondo interactions determines the ground state of these compounds. While the RKKY interaction favors a long-range magnetic order, the Kondo interactions have a tendency to screen the magnetic moments which lead to a non-magnetic ground state.\cite{Doniach, Lacroix} The result of these two interactions is summarized in the Doniach phase diagram.\cite{Doniach} A quantum critical point exists in the regime where the strength of RKKY and Kondo interactions are comparable. Quantum criticality in heavy fermion systems is one of the hot topics in condensed matter physics. In Ce-based compounds, quantum criticality and associated unusual physical properties such as the non-Fermi liquid behavior are found near the magnetic non-magnetic crossover, where the Ce valency is very close to $3+$. Usually one uses a tuning parameter such as pressure, doping or magnetic field to tune the electronic ground state of the system close to quantum critical point. Many Ce-based compounds exhibit magnetically mediated unconventional superconductivity in the quantum critical regime. Despite the first such superconductor, CeCu$_{2}$Si$_{2}$,\cite{Steglich} was discovered 35 years ago, many fundamental questions have yet not been settled and there is still an intensive research going on these superconductors. Pressure experiments on pure and Ge-doped CeCu$_{2}$Si$_{2}$ revealed two superconducting domes, one at lower pressure attributed to the magnetic quantum critical point and the other one at a higher pressure which could be related to valence fluctuation. \cite{Yuan, Lengyel} In contrast, unconventional superconductivity was not found in Yb-compounds near a magnetic quantum critical point, even though there are examples of Yb-compounds showing quantum criticality. YbAlB$_{4}$ is the only compound showing superconductivity which is proposed to be connected with critical valence fluctuations.\cite{Nakatsuji, Matsumoto, Yosuke} This provided additional incentive for the investigations of valence fluctuation phenomenon.

In our search for new valence fluctuating compound that may show quantum criticality similar to that of YbAlB$_{4}$, we have synthesized and investigated the physical properties of CeMo$_{2}$Si$_{2}$C . To the best of our knowledge, there is no report in the literature on the layered intermetallic compound CeMo$_{2}$Si$_{2}$C. The structurally homologous compound, PrMo$_{2}$Si$_{2}$C\cite{Dashjav} has been reported to form in CeCr$_{2}$Si$_{2}$C-type (``filled" CeMg$_{2}$Si$_{2}$-type) tetragonal crystal structure (space group \textit{P4/mmm}). The magnetic measurements on PrMo$_{2}$Si$_{2}$C have suggested a trivalent state of Pr in this compound. RCr$_{2}$Si$_{2}$C (R = Y, La-Sm, Gd-Er), an isoelectronic structural homologue series of CeMo$_{2}$Si$_{2}$C, has been reported to exhibit ferromagnetic ordering of rare earth ions\cite{Klosek} but intermediate valence for R = Ce.\cite{Tang, Mukherjee} We present the structural and physical properties of CeMo$_{2}$Si$_{2}$C, viz., x-ray diffraction (XRD), dc magnetic susceptibility, electrical resistivity, specific heat and x-ray absorption spectroscopy (XAS) and discuss the results in the context of the calculated electronic structure. From our measurements, we show that Ce ions are in a valence fluctuating state in this compound.
\section{Methods}
\label{sec:Methods}
\subsection{Experiment}
Polycrystalline sample of CeMo$_{2}$Si$_{2}$C has been prepared by the standard arc melting technique on a water cooled copper hearth. Since the melting point of Molybdenum element is very high, therefore, we first arc melted Si and Mo together in the stoichiometric ratio. Then the as-obtained pellet was arc melted with carbon and subsequently with cerium. All the constituent elements were of high purity (99.9\% and above). The samples were flipped after each melting and were melted several times to ensure homogeneity. The arc melted button was sealed in the Ta crucible and annealed in a vacuum furnace at 1500$^\circ$C for five days. The phase purity of the sample was checked by powder x-ray diffraction using Cu-$K_\alpha$ radiation and scanning electron micrograph (SEM). Energy dispersive x-ray (EDX) analysis was used to check the stoichiometry of the sample. Electrical resistivity measurements were carried out using conventional four probe technique in the temperature range 2-300~K in a physical property measurement system (PPMS, Quantum design). A commercial superconducting quantum interference device (SQUID) magnetometer (MPMS, Quantum design) was used for magnetic measurements. The specific heat was measured by relaxation method in a physical property measurement system (PPMS, Quantum design).  The absorption spectra of CeMo$_{2}$Si$_{2}$C at the Ce-L$_{III}$ edge (E = 5723~eV) have been recorded in transmission mode at the EXAFS beam line Al using the Si $(111)$ double crystal monochromator. The absorption behavior has been determined in the energy range 5450 eV to 6200~eV with a minimal step size of $\Delta E = 0.2$~eV close to the absorption edge e.g. $5723 \pm 25$~eV. The powdered sample with particle size smaller than $20  \mu$m was distributed on self adhesive Kapton foil. Multi-layers of 4 foils corresponding to 8 mg/cm${^2}$ resulted an absorption edge step of $\Delta \mu = 0.5$ which represents the difference of the low and high energy background functions at the absorption edge. Evaluation of spectra, energy calibration and normalization have been performed with the Athena modul of the Horae software package.\cite{Athena} The simultaneously measured reference spectra of the Ce$^{3+}$ compound CePO$_{4}$ was used for energy calibration and was measured together with the CeO$_{2}$ reference material with a dominant Ce$^{4+}$ contribution.
\subsection{Method of calculation}
We have performed the density-functional band structure calculations using two full potential codes: local orbital minimum basis band structure scheme (FPLO)\cite{Koepernik} and full-potential linearized augmented plane wave (FLAPW) method implemented in WIEN2K code.\cite{Blaha} Perdew-Burke-Ernzerhof (PBE) form of the generalized gradient approximation (GGA) was employed for the exchange correlation potential.\cite{Anis}Additionally, to account for the strong Coulomb repulsion within the Ce~$4f$ orbitals a typical value of the Coulomb energy U = 6~eV has been chosen for the GGA+U calculations.\cite{Perdew, Kaczorowski} The calculations were performed using the experimentally obtained lattice parameters of CeMo$_{2}$Si$_{2}$C where the full-lattice optimizations were done using Rietveld refined atomic coordinates.

\section{RESULTS AND DISCUSSION}
\label{sec:MD}
\subsection{Experiment}

\begin{figure}[htb!]
\centering
\includegraphics[width=10cm, keepaspectratio]{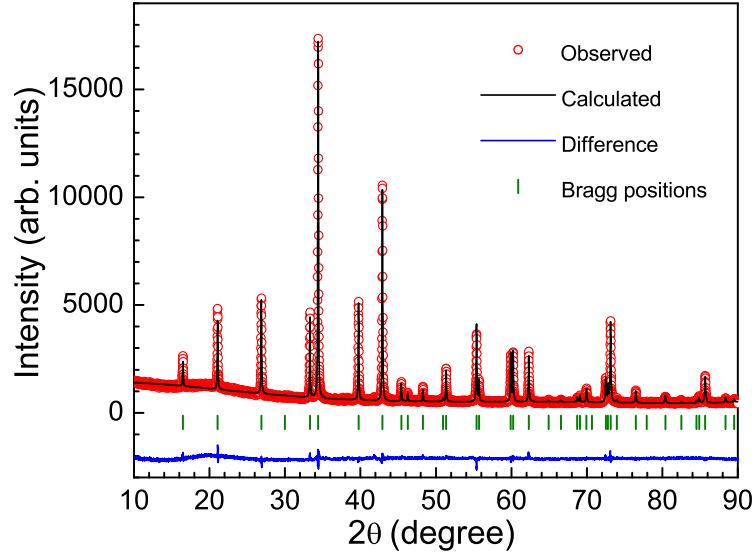}
\caption{\label{fig:CeMo2Si2C_XRD} (Color online) The powder x-ray diffraction pattern of CeMo$_{2}$Si$_{2}$C recorded at room temperature. The solid line through the experimental points is the Rietveld refinement profile calculated for the CeCr$_{2}$Si$_{2}$C-type layered tetragonal crystal structure (space group P4/mmm). The short vertical bars indicate the Bragg peak positions. The lowermost curve represents the difference between the experimental and model results.}
\end{figure}
Figure~\ref{fig:CeMo2Si2C_XRD} shows the XRD pattern of CeMo$_{2}$Si$_{2}$C together with the structural Rietveld refinement profile for the CeCr$_{2}$Si$_{2}$C-type layered tetragonal crystal structure (space group \textit{P4/mmm}). Powder x-ray diffraction data were collected on the crushed sample and analyzed by Rietveld refinement using {\tt FullProf} software.\cite{Rodriguez} The obtained lattice parameters for CeMo$_{2}$Si$_{2}$C are $a$ = 4.213(1)~{\AA}, and $c$ = 5.376(1)~{\AA} which are smaller than the values reported for homologous PrMo$_{2}$Si$_{2}$C ($a$ = 4.214~{\AA} and $c$ = 5.409~{\AA}).\cite{Dashjav} That means CeMo$_{2}$Si$_{2}$C unit cell volume does not follow the usual lanthanide contraction which is a first indication for a valence fluctuating Ce states.

\begin{figure}[htb!]
\centering
\includegraphics[width=8cm, keepaspectratio]{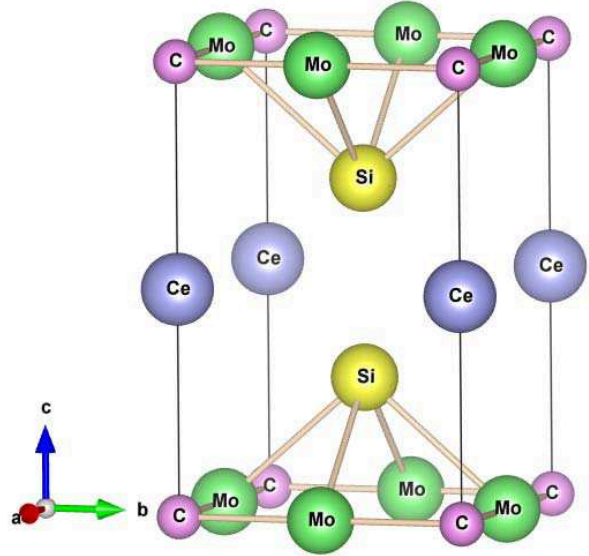}
\caption{\label{fig:unitcell}(Color online) Tetragonal crystal structure of CeMo$_{2}$Si$_{2}$C (space group \textit{P4/mmm}). The atomic coordinates are listed in Table I.}
\end{figure}

\begin{table}[ht]
\centering
\caption{\label{tab:XRD} Crystallographic parameters obtained from the Rietveld refinement of powder x-ray diffraction data for CeMo$_2$Si$_2$C.}
\begin{tabular}{l c c c c}
\hline
\hline
Structure &\multicolumn{4}{l} {CeCr$_{2}$Si$_{2}$C-type Tetragonal} \\
Space group & \textit{P4/mmm} \\
\multicolumn{3}{l}{Lattice parameters} \\
 $a$ (\AA)  & 4.213(1)  \\
 $c$ (\AA)  & 5.376(1)  \\
 $V_{cell}$ (\AA$^3$)& 95.42(1) \\
Refined Atomic Coordinates\\
\hline
Atom & \hspace{-5cm}Wyckoff &  \hspace{-2cm}x & ~~~~~y &  ~~~~z\\
\hline
~~Ce & \hspace{-5cm}$1b$ &  \hspace{-2cm}0.0000 & ~~~~~ 0.0000 &  ~~~~0.5000 \\
~~Mo & \hspace{-5cm}$2f$ &  \hspace{-2cm}0.0000 &  ~~~~~0.5000 &  ~~~~0.0000 \\
~~Si & \hspace{-5cm}$2h$ &  \hspace{-2cm}0.5000 &  ~~~~~0.5000 &  ~~~~0.2812 \\
~~C  & \hspace{-5cm}$1a$ &  \hspace{-2cm}0.0000 &  ~~~~~0.0000 &  ~~~~0.0000 \\
\hline
\hline
\end{tabular}
\end{table}

The crystallographic parameters obtained from the least square refinement of XRD data are listed in Table~\ref{tab:XRD}. The XRD and SEM reveal the single phase nature of the sample. The EDX composition analysis confirmed the desired stoichiometry of CeMo$_{2}$Si$_{2}$C. The CeCr$_{2}$Si$_{2}$C-type primitive tetragonal crystal structure of CeMo$_{2}$Si$_{2}$C is presented schematically in Fig.~\ref{fig:unitcell}. This structure is different from the very common tetragonal ThCr$_{2}$Si$_{2}$-type structure: an additional carbon atom sharing the same edge with Mo forms a layer of MoC. The loss of the body centering results in the $c/a$ ratio 1.27 which is almost half of usually observed ratio of about 2.5 in the ThCr$_{2}$Si$_{2}$-type structure. The value of the lattice parameter $c$ in this structure suggests a covalent bonding between the Ce-C along $[001]$ direction.\cite{Tang} The Ce atoms lie at the mid-point of the $c$-axis of the unit cell and there is no direct bonding between them.

\begin{figure}[htb!]
\centering
\includegraphics[width=10cm, keepaspectratio]{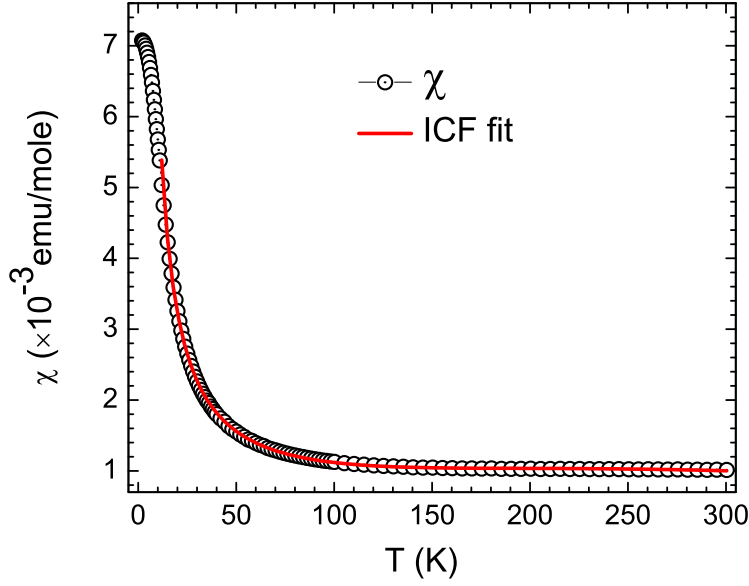}
\caption{\label{fig:susceptibility_ICF}(Color online) Temperature dependence of the magnetic susceptibility of CeMo$_{2}$Si$_{2}$C at 5~T. The solid red line through the magnetic susceptibility data shows a fit to the two-level ionic interconfiguration fluctuations (ICF) model in the temperature range 10-300~K.}
\end{figure}

The temperature dependence of magnetic susceptibility $\chi(T)$ of CeMo$_{2}$Si$_{2}$C measured under an applied magnetic field of $H = 5$~T is shown in Fig.~\ref{fig:susceptibility_ICF}. The very low value of susceptibility and its weak temperature dependence down to 100~K clearly indicate that Ce ions are in valence fluctuating state. Similar shape of susceptibility curves has been observed and are typical for Ce-based valence fluctuating compounds.\cite{Layek, Rojas, Kowalczyk, Mazumdar} Sometimes the magnetic susceptibility does show a curie tail at lower temperatures in valence fluctuating systems for instance in Ce$_{2}$Co$_{3}$Ge$_{5}$,\cite{Layek} Ce$_{2}$Ni$_{3}$Si$_{5}$,\cite{Mazumdar} and CeNi$_{2}$B$_{2}$C.\cite{Alleno} which arises due to free Ce$^{3+}$ magnetic impurity ions.

The dc magnetic susceptibility for a valence fluctuating system can be described with the two level ionic interconfiguration fluctuation (ICF) model. The theory of ICF was first proposed by Hirst \cite{Hirst} and latter Sales and Wohlleben developed it further to explain the valence fluctuating behavior observed in few Yb-based compounds. \cite{Sales} According to the ICF model,\cite{Franz} the temperature dependence of the susceptibility is given by
\begin{equation}
\chi(T) = \left(\frac{N}{3k_{B}}\right) \left[\frac{\mu_{n}^2 \nu(T)+ \mu_{n-1}^2\{1-\nu(T)\}}{T^*}\right]
\label{eq:chi1}
\end{equation}

with

\begin{equation}
\nu(T) = \frac{2J_{n}+1}{(2J_{n}+1) + (2J_{n-1}+1)exp(-E_{ex}/K_{B}T^*)}
\label{eq:chi2}
\end{equation}

and
\begin{equation}
T^* = \left[T^2+T_{sf}^2\right]^{1/2}
\label{eq:chi3}
\end{equation}

\noindent where $\mu_{n}$ and $\mu_{n-1}$ are the effective moments in $4f^n$ and $4f^{(n-1)}$  states, and $(2J_{n}+1)$ and $(2J_{n-1}+1)$ are the degeneracies of the corresponding energy states E$_n$ and E$_{n-1}$. Here E$_{ex}$ is interconfigurational excitation energy which is equals to (E$_{n}$-E$_{n-1})$ and $T_{sf}$  is the spin fluctuation temperature associated with the valence fluctuation. In order to take care of the contribution of Ce$^{3+}$ ions which belong to the impurity phases present in CeMo$_{2}$Si$_{2}$C, we have added the term $\chi_{imp} = n*C/(T-\theta)$ to the magnetic susceptibility. Thus the final equation is
\begin{equation}\begin{split}
\chi(T)& = (1-n)\left(\frac{N}{3k_{B}}\right)\left[\frac{(2.54\mu_{B})^2\{1-\nu(T)\}}{T^*}\right]\\
\quad & + n\frac{C}{T-\theta}+\chi_{0}
\label{eq:chi4}
\end{split}\end{equation}

with

\begin{equation}
\nu(T) = \frac{ 1}{1 + 6exp(-E_{ex}/K_{B}T^*)}
\label{eq:chi5}
\end{equation}

\noindent where $\chi_{0}$ is temperature independent term and n is the fraction of stable Ce$^{3+}$ ions. Here we have taken Ce$^{4+}$ (J = 0 and $\mu$ = 0~$\mu_{B}$) state as ground state and Ce$^{3+}$ (J = $\frac{5}{2}$ and $\mu$ = 2.54~$\mu_{B}$) as an excited state. The solid line through the magnetic susceptibility data in Fig.~\ref{fig:susceptibility_ICF} is the fit to the equation 4. The values of parameters obtained from the least square fits are $n$ = 0.06, E$_{ex}$ = 677 $\pm$ 12~K, $T_{sf}$ = 205 $\pm$ 3~K, $\theta$ = 1.9~K   and $\chi_{0}= -1.67\times 10^{-4}$ emu/mole. The values of E$_{ex}$ and $T_{sf}$ are in the range of typical values found for the valence-fluctuating system Ce$_{2}$Rh$_{3}$Si$_{5}$ (E$_{ex}$ = 845 K and $T_{sf} = 129$~K).\cite{Kaczorowski}
\begin{figure}[htb!]
\centering
\includegraphics[width=10cm, keepaspectratio]{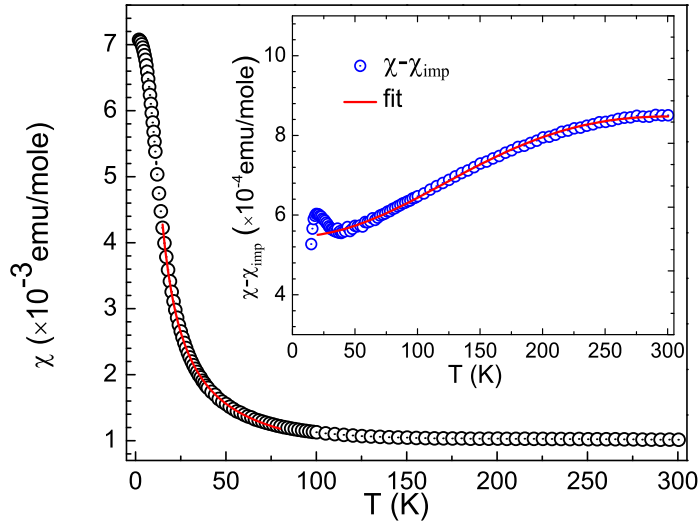}
\caption{\label{fig:susceptibility_imp}(Color online) Temperature dependence of the magnetic susceptibility of CeMo$_{2}$Si$_{2}$C at 5~T. The solid line through the magnetic susceptibility data in the temperature range 15-80~K shows a fit to the equation $\chi(T)= \chi_{0}+ n*C/(T - \theta)$. The blue circles in the inset represent the susceptibility data obtained after subtracting the extrapolated impurity component from the measured susceptibility data in the temperaure range 10-300~K and the solid line through the data points represents an ICF fit as discussed in the text.}
\end{figure}

Furthermore, to understand the intrinsic behavior of the susceptibility for the sample i.e. $\chi(T)$ without the contribution from Ce$^{3+}$ impurity phase, we have subtracted the magnetic impurity contribution from the measured susceptibility data. The susceptibility data were first fitted to the equation $\chi(T)= \chi_{0}+ n*C/(T - \theta)$ in the temperature range 15-80~K where the dominant magnetic contribution is coming from Ce$^{3+}$ magnetic impurity pase (see Fig. 4). The parameters obtained from the least square fit, n = 0.06 and $\theta$ = 2.32~K are nearly same as found in the impurity part of ICF fitting discussed earlier. In next step, the impurity part $\chi_{imp}$ = $n*C/(T - \theta)$ was subtracted from the measured $\chi(T)$ in the temperature range 10-300 K. The resulted (after subtraction) susceptibility  $\chi-\chi_{imp}$ posses a broad maximum which is a characteristic feature of a valence fluctuating system\cite{Kaczorowski, Malik} (see inset of Fig. 4). The $\chi-\chi_{imp}$ data were fitted to the following ICF equation
\begin{equation}\begin{split}
\chi(T)& = (1-n)\left(\frac{N}{3k_{B}}\right)\left[\frac{(2.54\mu_{B})^2\{1-\nu(T)\}}{T^*}\right]\\
\quad & +\chi_{0}
\label{eq:chi4}
\end{split}\end{equation}
\noindent where the impurity concentration n is kept fixed at 0.06. The fit to the data yields the values E$_{ex}$ = 701 $\pm$ 4~K, $T_{sf}$ = 217 $\pm$ 1~K and $\chi_{0}= -1.24\times 10^{-4}$~emu/mole, which are close to the values obtained from the ICF fitting earlier above.
\begin{figure}[htb!]
\centering
\includegraphics[width=10cm, keepaspectratio]{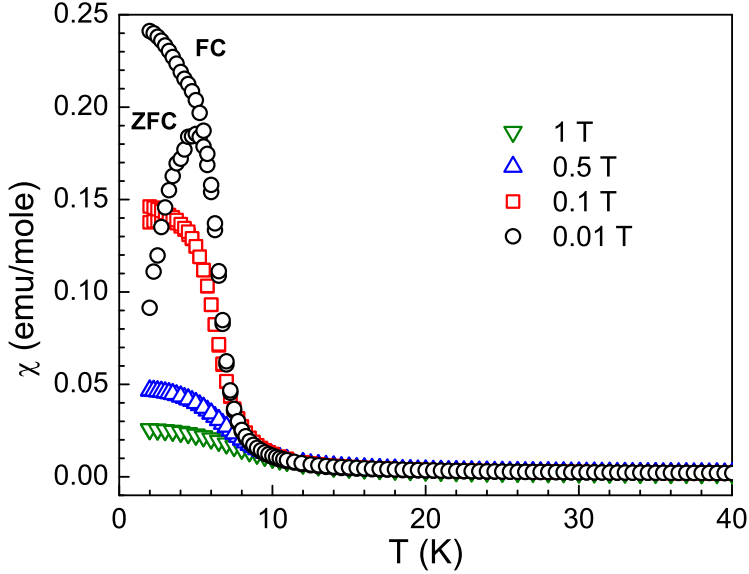}
\caption{\label{fig:susceptibility_fields}(Color online) The temperature dependence of zero field-cooled (ZFC) and field-cooled (FC) dc magnetic susceptibility in the temperature range 2-40~K measured under various applied magnetic fields.}
\end{figure}

Even though the valence fluctuating systems are not expected to order magnetically, the magnetic susceptibility measured under low magnetic fields exhibit sharp increase below 8 K (Fig.~\ref{fig:susceptibility_fields}). However, the low value of magnetization at 5~K $(\sim 0.003~\mu_{B}$/Ce at 0.01 T) and the absence of any anomaly in the specific heat data suggest that this low temperature magnetic susceptibility anomaly is not intrinsic and arises from small amount of impurity phase. The field dependence indicate this impurity phase to be ferromagnetic. Therefore it is likely a small amount of CeSi$_{2-x}$.\cite{Yashima}
\begin{figure}[htb!]
\centering
\includegraphics[width=10cm, keepaspectratio]{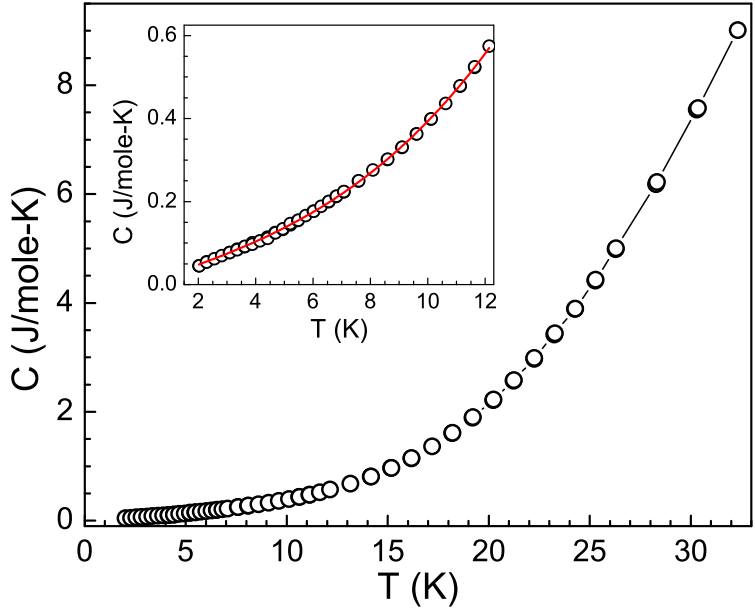}
\caption{\label{fig:SP_heat} (Color online) Temperature dependence of the specific heat data, $C(T)$ of CeMo$_{2}$Si$_{2}$C measured in zero field in the temperature range (1.8-35~K). The solid line through the specific heat data in the inset shows the fit to the equation $C(T) = \gamma T + \beta T^{3}$.}
\end{figure}

The specific heat $C(T)$ of CeMo$_{2}$Si$_{2}$C measured under zero applied magnetic field is presented in Fig.~\ref{fig:SP_heat} in the temperature range 1.8-35~K. The specific heat exhibits no anomaly in this temperature range down to 1.8 K and hence confirms the absence of magnetic ordering in CeMo$_{2}$Si$_{2}$C. We could fit the low temperature specific heat data below 12~K to the equation
\begin{equation}
C(T) = \gamma T + \beta T^{3}
\label{eq:C}
\end{equation}
\noindent where $\gamma T$ is the electronic contribution to the specific heat, $\beta T^{3}$ is the phononic contribution to the specific heat (inset of the Fig.~\ref{fig:SP_heat}). The fit to the data yields the value of $\gamma = 23.4$~mJ/mol~K$^{2}$ and $\beta =1.6\times10^{-4}$~J/mol~K$^{4}$\@. This gamma value is at the lower border of the range expected for valence fluctuating Ce system, but comparable to the value found e.g. in CeRu$_{2}$ ($\gamma = 23$~mJ/mol~K$^{2}$),\cite{Allen} Ce$_{2}$Co$_{3}$Ge$_{5}$ ($\gamma = 17$~mJ/mol~K$^{2}$)\cite{Layek} and CeRu$_{3}$Si$_{2}$ ($\gamma = 39$~mJ/mol~K$^{2}$)\cite{Rauchschwalbe}. The Debye temperature $\Theta_{D}$ for CeMo$_{2}$Si$_{2}$C was estimated from $\beta$ using the relation\cite{Kittel}
\begin{equation}
\Theta_{D} = \left( \frac{12 \pi^{4} n R}{5 \beta} \right)^{1/3}
 \label{eq:Debye-Temp}
\end{equation}

\noindent where $R$ is molar gas constant and $n = 6$ is the number of atoms per formula unit (f.u.)\@. Therefore, we obtain Debye temperature $\Theta_{D}$ = 417~K\@ for CeMo$_2$Si$_2$C.

\begin{figure}[htb!]
\centering
\includegraphics[width=10cm, keepaspectratio]{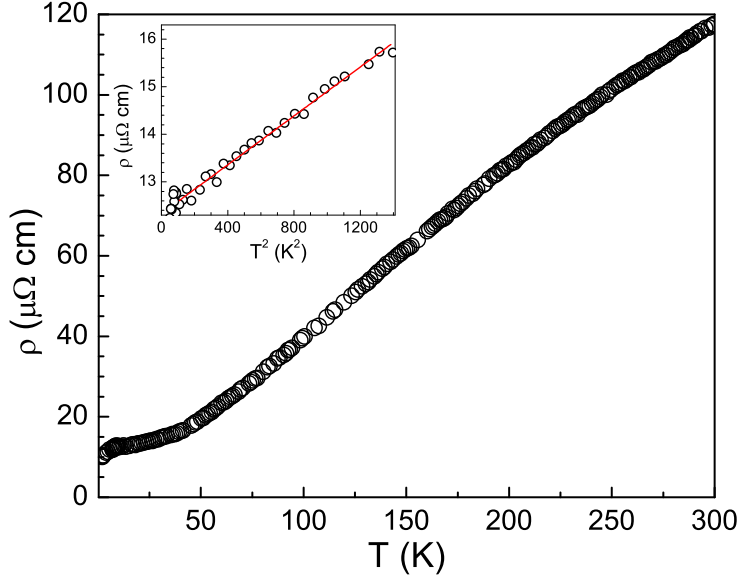}
\caption{\label{fig:Resistivity} (Color online) Temperature dependence of electrical resistivity data, $\rho(T)$ of CeMo$_2$Si$_2$C as a function of temperature. The solid line in the inset shows the fit to the equation $\rho(T) = \rho_{0}$ + $AT^{2}$.}
\end{figure}

The temperature dependence of electrical resistivity $\rho(T)$ of CeMo$_{2}$Si$_{2}$C is shown in Fig.~\ref{fig:Resistivity}. The electrical resistivity data exhibit metallic behavior with residual resistivity, $\rho_0 \sim 12.7 ~\mu \Omega$~cm at 2~K. The large value of residual resistivity ratio (RRR = $\rho_{300\,{K}}/\rho_{2\,{K}}) \sim  9$ indicates good crystallinity of our sample. We do not observe any prominent anomaly in the resistivity data and the data fit well to the following equation in the low temperature range 10-35~K , i.e.
\begin{equation}
\rho(T) = \rho_{0} + AT^2
\label{eq:Resistivity}
\end{equation}
The fitting yield the parameters $\rho_0$$\sim$12.3~$\mu~\Omega$~cm for the residual resistivity and the coefficient $A = 2.57 \times 10^{-3}~\mu~\Omega$~cm~K$^{-2}$. Thus, the electrical resistivity of CeMo$_{2}$Si$_{2}$C exhibits the Fermi liquid behavior as usually observed for Ce-based valence fluctuating systems.\cite{Andres} Within the Fermi liquid theory the $A$ coefficient is related to $\gamma^2$. Kadowaki and Woods found that in heavy fermion systems and valence fluctuating systems the ratio $A/\gamma^{2}$ is of the order of $1\times 10^{-5}~\mu~\Omega$~cm~(mol~K/mJ)$^{2}$.\cite{Kadowaki} Later on N. Tsujii et al.\cite{Tsujii} suggested this ratio to scale with 2/N(N-1) where N is the degeneracy of the orbital state of the $f$ element. For an intermediate valent Ce system N = 6 gives $A/\gamma^{2}$ = $6.7\times 10^{-7}~\mu~\Omega$~cm~(mol~K/mJ)$^{2}$. For CeMo$_{2}$Si$_{2}$C our experimental data result in $A/\gamma^{2} = 0.5\times 10^{-5}~\mu~\Omega$~cm~(mol~K/mJ)$^{2}$, just the original Kadowaki Wood value. Furthermore, in a Fermi liquid one expect a $T$-independent susceptibility $\chi_{FL}$ at low temperatures with a value scaling with $\gamma$ too. This is expressed in the Wilson Sommerfeld ratio i.e.
\begin{equation}
R_{W} = \frac{\pi^{2}k_{B}^{2} \chi_{FL}}{\gamma \mu_{eff}^{2}}
 \label{eq:RW}
\end{equation}
\noindent which is expected to be close to one. Taking the low temperature value of $\chi$$_{FL}$$ \approx 5.6\times 10^{-4}$~emu/mole from the impurity corrected $\chi(T)$ shown in the inset of Fig.4, $\gamma = 23.4$~mJ/mol~K$^{2}$ as determined above, and $\mu_{eff}$ = 2.54~$\mu_{B}$ as expected for the free Ce$^{3+}$ moment we get R$_{W}$ = 0.81, while taking $\mu_{eff}$ = 1.73 as for a free conduction electron, one get R$_{W}$ = 1.7. Thus, low $T$ susceptibility, specific heat and resistivity data fulfill Fermi liquid expectations.

\begin{figure}[htb!]
\centering
\includegraphics[width=10cm, keepaspectratio]{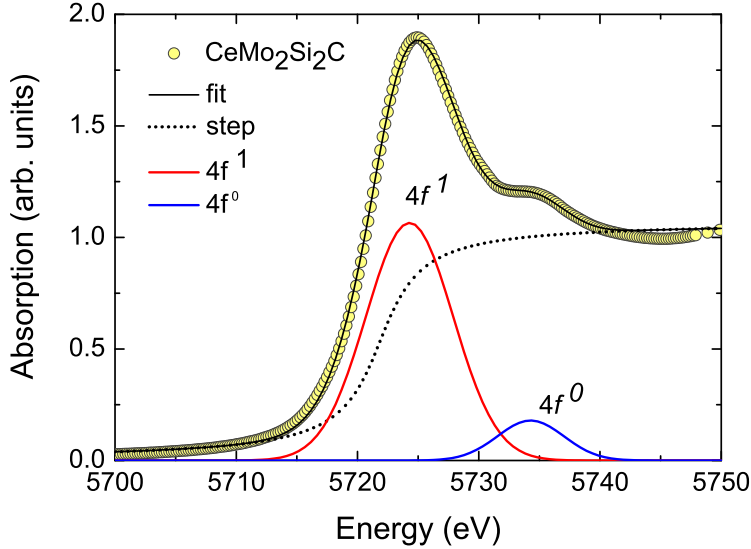}
\caption{\label{fig:XAS} (Color online) $L_{III}$-edge x-ray absorption of CeMo$_{2}$Si$_{2}$C at 300 K. The solid black line through data points shows the least square fit. The contributions due to $4f^1$ and $4f^0$ configurations are represented by the solid Gaussian lines, while the dashed line represents the arc tan function.}
\end{figure}

The Ce-$L_{III}$-edge x-ray absorption spectrum of CeMo$_{2}$Si$_{2}$C taken at room temperature is presented in Fig.~\ref{fig:XAS}. To analyze the spectrum in detail two Gaussian functions were taken into consideration which represents the $4f^1$ and $4f^0$ configuration. Additionally, an arc tan function was added to provide the relative weight of the two electron configurations. The larger peak at 5724~eV corresponds to the $4f^1$ configuration and the smaller peak at 5734~eV corresponds to the $4f^0$ configuration. The Ce-$L_{III}$ clearly indicates that the Ce ions in this compound are in valence fluctuating state. Moreover, the average formal $L_{III}$ valence of Ce can be obtained from the intensity ratio of the two Gaussian peaks using the equation $<$$\widetilde{\nu}$$>$  = 3+I$_{1}$/(I$_{1}$+I$_{2}$). Hence we get the value of average formal $L_{III}$ valence of Ce atoms in CeMo$_{2}$Si$_{2}$C as $<$$\widetilde{\nu}$$>$ = 3.11 at room temperature. This value is less than the value ($\sim$ 3.4) obtained for a tetravalent CeO$_{2}$.\cite{Vainshtein} Similar values of mean formal $L_{III}$ valence of Ce ions have also been reported for other valence fluctuating compounds, for example, CeRhSi$_{2}$ ($<$$\widetilde{\nu}$$>$ = 3.15), Ce$_{2}$Rh$_{3}$Si$_{5}$ ($<$$\widetilde{\nu}$$>$ = 3.18)\cite{Kaczorowski} and  CeNi$_{2}$B$_{2}$C ($<$$\widetilde{\nu}$$>$ = 3.18).\cite{Alleno} at room temperature.

\begin{figure}[htb!]
\centering
\includegraphics[width=8.5cm, keepaspectratio]{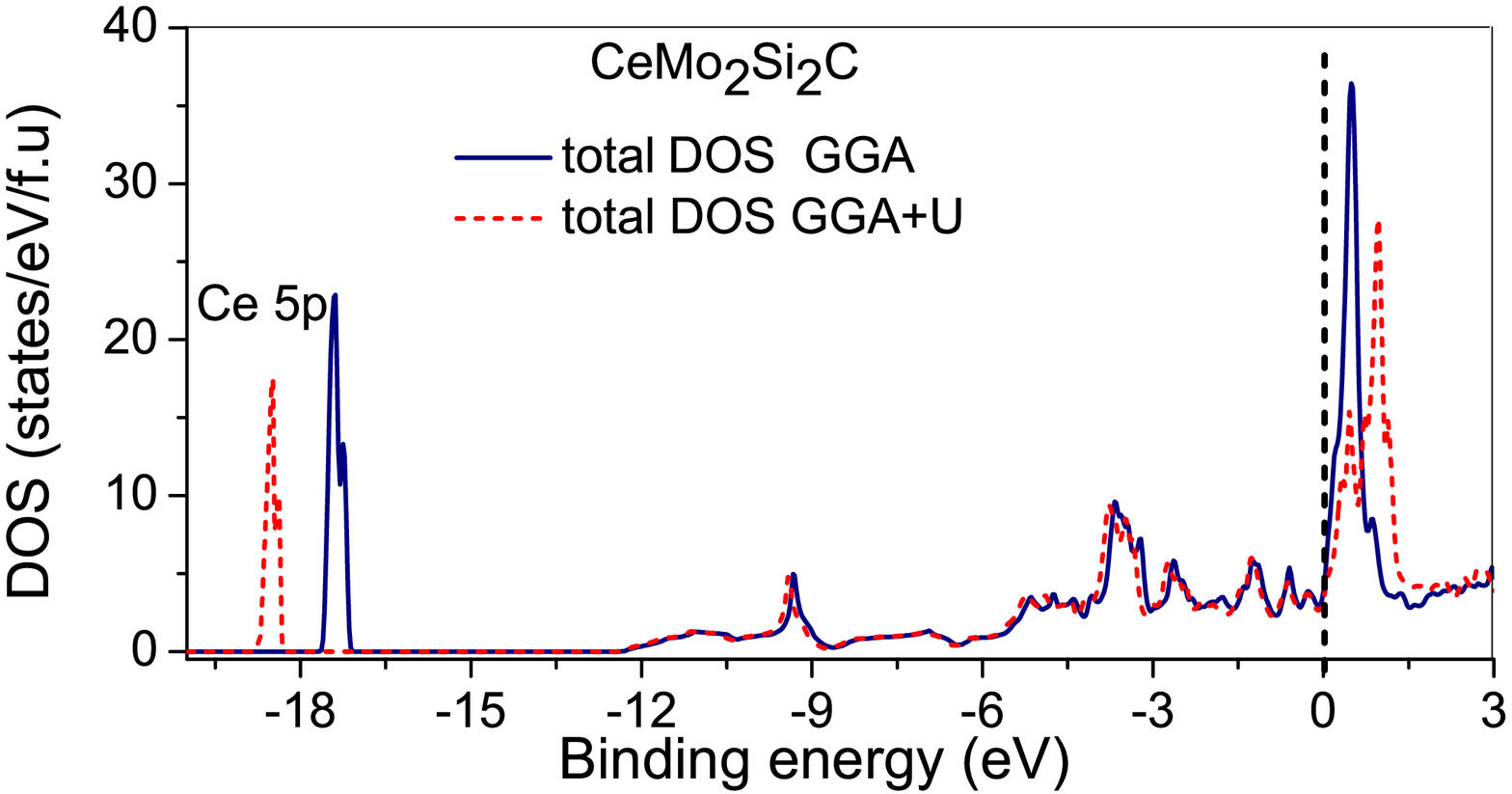}
\includegraphics[width=8.5cm, keepaspectratio]{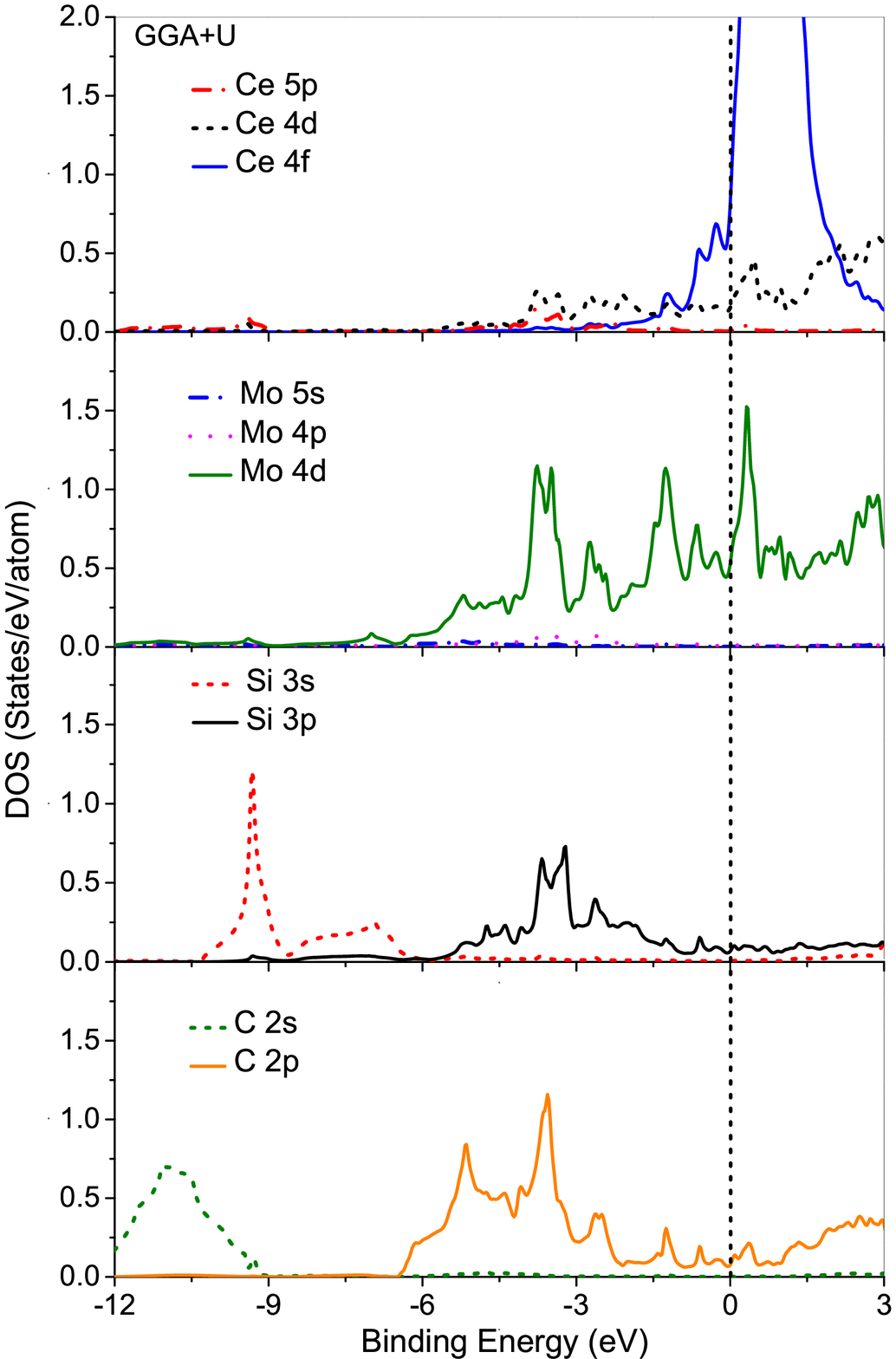}
\caption{\label{fig:dos} (Color online) Total (upper panel) and partial densities of states (bottom panels) for CeMo$_{2}$Si$_{2}$C. Here the Fermi level (E$_{F}$) corresponds to zero binding energy.}
\end{figure}

\subsection{Band Structure Calculations}
In order to gain some information on the electronic state, electronic structure calculations were performed for the compound using two different approaches, i.e., with FPLO and full potential LAPW method. Both the methods give very similar results and therefore only those obtained by using the FLAPW method are discussed here. Because of the strong correlations in the $4f$ shell, the density functional approach is not able to treat the $4f$-electrons of Ce accurately. In contrast non-$f$ valence states are usually only weakly affected by these correlations, except close to the Fermi level where hybridization with $f$-states becomes relevant. Therefore, such a calculation allows insight into the non-$f$ valence states and gives some information on the states which are hybridized with the Ce~$4f$ states at the Fermi level. In this method the unit cell is divided into non-overlapping muffin-tin spheres centered at the atomic sites and an interstitial region. The muffin-tin radii were chosen to be of 2.5~a.u. for Ce, 2.1~a.u. for Mo, 2.2~a.u. for Si and 1.86~a.u for C. The set of plane-wave expansion K$_{MAX}$ was determined as R$_{MT}\times$K$_{MAX}$ equal to 7 and K mesh used was $13\times13\times10$. We have performed the calculations with and without spin polarization. Notably the self-consistent results for magnetic solutions were nearly the same as those for non-magnetic ones because the spin polarized calculation resulted in a non-magnetic state. Total and local magnetic moments were approximately zero ($\sim10^{-5}\,\mu_{B}$/Ce) in the spin polarized calculations.

Figure~\ref{fig:dos} shows the total and partial Density of States (DOSs) for CeMo$_{2}$Si$_{2}$C, calculated within the GGA and GGA+U (U = 6~eV) approximations. To describe in detail, here we divide the total DOS into four main parts.
\begin{enumerate}
  \item The peak located below -16~eV is due to Ce~$5p$ electronic state. It is remarkable that the GGA+U peak is shifted towards higher binding energy by 1~eV with respect to GGA.
  \item The sub-band situated between -12~eV and -6~eV, is formed mainly by Si~$3s$ and C~$2s$ states.
  \item The third part, which is located between the -6 to -2~eV contains strongly hybridized Mo~$4d$ , Si~$3p$ and C~$2p$ states. Hybridized states of Ce~4d and C~2p are observed as well just below the Fermi level (see the partial density of states of Fig.~\ref{fig:dos}).
  \item The bottom of the conduction band or at the Fermi level, the main contribution to the DOS is provided by Ce~$4f$ and Mo~$4d$ electronic states. Hybridization between Ce~$4f$ and Mo~$4d$ states are observed near the Fermi level. In case of GGA+U the main peak (Ce~$4f$) is broadened and slightly shifted towards right side above the Fermi level in comparison to GGA and consequently reduced the occupancy of Ce~$4f$ electrons at the Fermi level.
\end{enumerate}
It should be noted that at the Fermi level a small contribution to the total DOS come from Ce~$5d$, Si~$3p$ and C~$2p$ states as well. The orbital occupancy of electrons inside the atomic sphere of the Ce atom is calculated by the GGA approach as $6s^{1.98}$ $5d^{0.64}$ $4f^{0.94}$ and by GGA+U calculation as $6s^{1.99}$ $5d^{0.85}$ $4f^{0.64}$. Thus the f-count in these calculations is slightly (GGA) or far (GGA+U) below 1. Comparison with the $LIII$ result shows that GGA overestimate the $f$ count while GGA+U provides more realistic value of $f$ electron count.

\begin{table}[ht]
\centering
\caption{\label{tab:dos} Total densities of states $N[E_{F}]$ (in states/eV/f.u) and partial densities of states (in states/eV/atom) at the Fermi level, the calculated Sommerfeld coefficient $\gamma_{0}$ (in ~mJ/mol~K$^{2}$) and molar Pauli paramagnetic susceptibility $\chi_{FL}$ (in $10^{-4}$~emu/mol).}
\setlength{\tabcolsep}{6pt}
\begin{tabular}{c c c c c c}
\hline
\hline
Method &Total $N[E_{F}]$  &  $\gamma_{0}$  &  $\chi$ & Ce-$4f$ &  Mo-$4d$ \\
\hline
GGA     &3.86  &  9.10  &  1.24  &  1.47  &  0.52  \\ [1ex]
GGA+U   &3.16  &  7.45  &  1.01  &  0.77  &  0.52  \\ [0.5ex]
\hline
\hline
\end{tabular}
\end{table}

Based on the DOS at the Fermi level here we obtained the Sommerfeld coefficients $\gamma_{0}$ and the Pauli paramagnetic susceptibility $\chi_{FL}$ as calculated under the assumption of free electron model, $\gamma_{0}= (\pi^{2}/3)N[E_{F}]K_{B}^2$ and $\chi_{FL} = \mu_{B}^{2}N[E_{F}]$, where $N[E_{F}]$  is the electron density of states at the Fermi level. All the calculated values are listed in Table~\ref{tab:dos}. The calculated $\gamma_{0}$ and $\chi_{FL}$ values are a factor of about 3 and 6, respectively, below the experimental values. This difference can be very simply attributed to the strong renormalization connected with the valence fluctuations.

\begin{figure}[htb!]
\centering
\includegraphics[width=10cm, keepaspectratio]{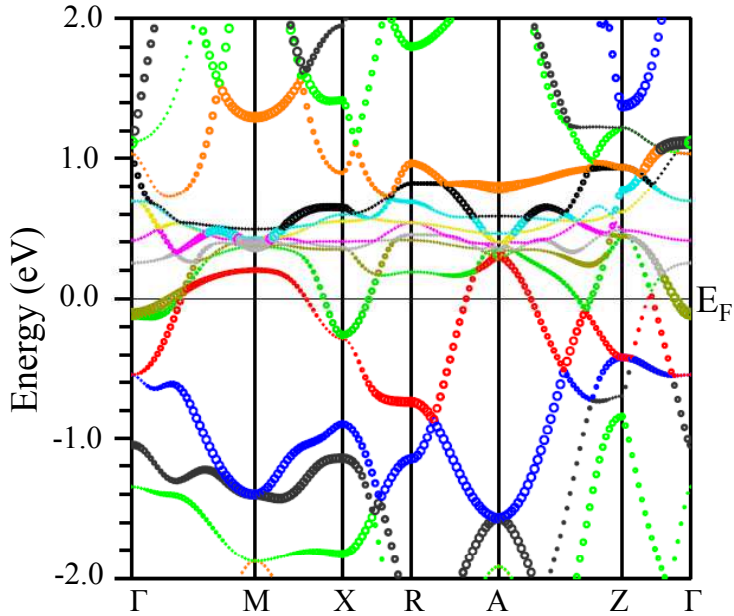}
\caption{\label{fig:bandstructure} (Color online) Band structure of CeMo$_{2}$Si$_{2}$C along the symmetry lines of the Primitive tetragonal BZ. Mo~$4d$ contribution to the bands is represented by the larger circles. Different colors of the bands are used just to guide the eyes.}
\end{figure}
The band structure along various symmetry directions calculated within the GGA is shown in Fig.~\ref{fig:bandstructure}. The band structure near the Fermi level is mainly composed of three bands which are crossing E$_{F}$. The fat part (bigger circles) of the bands represent the contribution from Mo $4d$ states which are considerably hybridized with the Ce $4f$ bands near E$_{F}$. Two hole like bands cut by the Fermi level near $\Gamma$ and $X$ points and a highly dispersed electron like band crosses the Fermi level near $M$ and $A$ points. The strong dispersion along $\Gamma$-$Z$ shows that this compound, despite having a quasi 2D structure, has a true 3D electronic structure. Interesting features are observed between $\Gamma$-$Z$ and $A$-$Z$ as the sharp tip of the bottom of the conduction band touches the sharp tip of the top of the valence band.

\section{Conclusion}
\label{sec-Conclusion}
To summarize, our comprehensive studies of physical properties of CeMo$_{2}$Si$_{2}$C by $\chi(T)$, $C(T)$, $\rho(T)$, and XAS measurements reveal the valence fluctuating state of Ce in CeMo$_{2}$Si$_{2}$C. The compound crystallizes in the CeCr$_{2}$Si$_{2}$C-type layered tetragonal structure (space group \textit{P4/mmm}). The value of magnetic susceptibility is very low and almost independent of temperature except a curie tail at lower temperature (as observed in many valence fluctuating compounds) which arises due to a small amount ($\sim$ 6 \%) of an impurity Ce$^{3+}$ phase present in the sample. An ICF analysis of $\chi(T)$ result in a fluctuation temperature $T_{sf}$ = 205 $\pm$ 3~K and an interconfigurational excitation energy E$_{ex}$ = 677 $\pm$ 12~K. The temperature dependence of specific heat at zero magnetic field does not show any anomaly throughout the temperature range and hence confirms the absence of any magnetic ordering down to 2~K. The Sommerfeld coefficient obtained from the specific heat data $\gamma$ = 23.4~mJ/mol~K$^{2}$ is relatively low for a valence fluctuating compound which indicates a strong hybridization of Ce~$4f$ states with the conduction states at the Fermi level. The electrical resistivity exhibits a metallic behavior with a $T^{2}$ dependence in the low temperature range indicating the Fermi liquid behavior as observed for many Ce-based valence fluctuating systems. XAS studies indicate an average formal $L_{III}$ valence $<$$\widetilde{\nu}$$>$ = 3.11 hence gives a direct evidence for valence fluctuations. Nevertheless, this value is less than that obtained for a tetravalent CeO$_{2}$ ($<$$\widetilde{\nu}$$>$ $\simeq$ 3.4). The density functional calculations show strong dispersions along all direction evidencing a 3D electronic structure despite the quasi 2D chemical structure. Strong hybridization is found between Mo~$4d$, Si~$3p$ and C~$2p$ states, and between Ce~4d and C~2p states at lower binding energies below the Fermi level. The bands at the Fermi level bear a large Mo~$4d$ character and provide the states which are hybridized with the Ce $4f$ states.
\ack{}
This work has been supported by the Council of Scientific and Industrial Research, New Delhi (Grant No. 80(0080)/12/ EMR-II).

\end{document}